\documentclass[reprint,amsmath,amssymb,aps,prm,superscriptaddress,showkeys]{revtex4-2}
\usepackage{graphicx}               
\usepackage[colorlinks=true,linkcolor=blue,citecolor=blue,urlcolor=black]{hyperref} 
\usepackage{xcolor}
\usepackage{amsmath,amssymb}

\makeatletter

\newcommand{\Rmnum}[1]{\expandafter\@slowromancap\romannumeral #1@}
\makeatother

\newcommand{\BOS}{Bi$_2$O$_2$Se}
\newcommand{\CVS}{cm$^2$V$^{-1}$s$^{-1}$}
\newcommand{\STO}{SrTiO$_3$}
\usepackage{amstext}

\begin{document}

\title{Polar discontinuities and interfacial electronic properties of Bi$_2$O$_2$Se on SrTiO$_3$}

\author{Ziye Zhu}
\affiliation{School of Materials Science and Engineering, Zhejiang
  University, Hangzhou 310027, China}
\affiliation{Key Laboratory of 3D Micro/Nano Fabrication and
  Characterization of Zhejiang Province, School of Engineering,
  Westlake University, Hangzhou 310030, China}
\affiliation{Research Center for Industries of the Future, Westlake
  University, Hangzhou 310030, China}

\author{Jingshan Qi}
\affiliation{School of Science, Tianjin University of Technology,
  Tianjin 300384, China}

\author{Xiaorui Zheng}
\affiliation{Key Laboratory of 3D Micro/Nano Fabrication and
  Characterization of Zhejiang Province, School of Engineering,
  Westlake University, Hangzhou 310030, China}
\affiliation{Research Center for Industries of the Future, Westlake University, Hangzhou 310030, China}

\author{Xiao Lin}
\affiliation{Key Laboratory for Quantum Materials of Zhejiang Province, School of Science, Westlake University, Hangzhou 310030, China}

\author{Wenbin Li}
\email{liwenbin@westlake.edu.cn}
\affiliation{Key Laboratory of 3D Micro/Nano Fabrication and
  Characterization of Zhejiang Province, School of Engineering,
  Westlake University, Hangzhou 310030, China}
\affiliation{Research Center for Industries of the Future, Westlake University, Hangzhou 310030, China}

\date{\today}


\begin{abstract}
The layered oxychalcogenide semiconductor \BOS\ (BOS) hosts a
multitude of unusual properties including high electron
mobility. Owing to similar crystal symmetry and lattice constants, the
perovskite oxide \STO\ (STO) has been demonstrated to be an excellent
substrate for wafer-scale growth of atomically thin BOS
films. However, the structural and electronic properties of the
BOS/STO interface remain poorly understood. Here, through
first-principles study, we reveal that polar discontinuities and
interfacial contact configurations have a strong impact on the
electronic properties of ideal BOS/STO interfaces. The lowest-energy
\mbox{[Bi-TiO$_2$]} contact type, which features the contact between a
Bi$_2$O$_2$ layer of BOS with the TiO$_2$-terminated surface of STO,
incurs significant interfacial charge transfer from BOS to
STO, producing a BOS/STO-mixed, $n$-type metallic state at the
interface. By contrast, the \mbox{[Se-SrO]} contact type, which is the
most stable contact configuration between BOS and SrO-terminated STO
substrate, has a much smaller interfacial charge transfer from STO to
BOS and exhibits $p$-type electronic structure with much weaker
interfacial hybridization between BOS and STO. These results indicate
that BOS grown on TiO$_2$-terminated STO substrates could be a fruitful system for exploring emergent phenomena at the interface between an oxychalcogenide and an oxide, whereas BOS grown on SrO-terminated substrates may be more advantageous for
preserving the excellent intrinsic transport properties of BOS.
\end{abstract}


\maketitle

\section{Introduction}

Bismuth oxyselenide \BOS\ (BOS) has recently emerged as a layered
semiconductor with a moderate bandgap ($E_g\sim0.8$~eV), excellent air
stability, ultrahigh electron mobility, and extraordinary optical
sensitivity~\cite{Wu2017n,Chen2018,Yin2018,Li2021AMR}. Room-temperature
electron Hall mobility as high as 450~cm$^2$V$^{-1}$s$^{-1}$ has been
measured in BOS ultrathin films (thickness $\sim$6~nm) grown by
chemical vapor deposition (CVD)~\cite{Wu2017n}.  At low temperatures,
the electron mobility of such BOS thin films can reach a huge number
above 20,000~cm$^2$V$^{-1}$s$^{-1}$~\cite{Wu2017n}, rivalling that of
two-dimensional (2D) electron gas at the LaAlO$_3$/SrTiO$_3$ (LAO/STO)
interface, as well as graphene samples grown by chemical vapor
deposition (CVD)~\cite{Ohtomo2004, Petrone2012}. A huge static
dielectric constant ($\epsilon_0>$150), which results from the
proximity to a ferroelectric transition and strongly suppressing
Coulombic defect scattering, is crucial for the observed ultrahigh
electron mobility~\cite{Zhu2022,Xu2021}. Moreover, high-quality,
stable native oxide dielectric Bi$_2$SeO$_5$ can directly form on top
of BOS via layer-by-layer oxidization while preserving the high
electron mobility of BOS, presenting a unique advantage of BOS over
other 2D materials in terms of compatibility with existing
silicon-based semiconductor technology~\cite{Li2020NE}.

For commercial success, functional BOS devices need to be produced at
large scale and with uniform characteristics. Important in this
respect, it has recently been shown that high-quality,
single-crystalline BOS thin films and atomic layers can be grown on
STO substrates at wafer scale using CVD or molecular beam
epitaxy (MBE)~\cite{Tan2019, Liang2019}, owing to the symmetry and
lattice matching between BOS and STO. However, the measured electron
Hall mobility of BOS thin films grown on STO substrates
($\sim$94~\CVS) was found to be poorer than those grown on mica
substrates (200--450~\CVS)~\cite{Tan2019}. Although interfacial
scattering was proposed as a possible explanation of the mobility
degradation, the exact microscopic origin remains
unclear. Furthermore, unlike conventional layered semiconductors such
as MoS$_2$, whose layers are bound together by van der Waals (vdW)
interaction, BOS features electrostatic interaction between positively
charged Bi$_2$O$_2$ layers and negatively charged Se layers. This
feature could lead to stronger interfacial bonding and interaction
between BOS and STO than those between a conventional vdW layered
semiconductor and STO, resulting in richer interfacial
phenomena. Indeed, the close symmetry and lattice-constant matching
between BOS and STO, as well as between BOS and other
perovskite-related materials~\cite{Zhu2022}, could enable the growth
of a wide range of BOS-based artificial heterostructures with emergent
properties. In the past, exotic phenomena such as interface
superconductivity, strong electro-magnetic coupling, and fractional
quantum Hall effect have been observed at the interfaces of two
perovskite oxides such as between LAO and
STO~\cite{Hwang2012}. By contrast, little has been explored with
respect to the interfacial properties between an oxychalcogenide (to
which BOS belongs) and a perovskite oxide.

These attractive prospects have motivated us to investigate the
structural and electronic properties of ideal BOS/STO interfaces
via first-principles calculations. The results of our study reveal
that interfacial contact configurations have a strong influence on the
electronic properties of the BOS/STO interfaces, which originates from
the discontinuity of polarity (``polar discontinuity'') at the
interface between BOS and STO. As a result of the polar discontinuity
and the subsequent electronic reconstruction, the lowest-energy
\mbox{[Bi-TiO$_2$]} interface, formed between a Bi$_2$O$_2$ layer of
BOS and a TiO$_2$-terminated surface of STO, features a significant
amount of interfacial charge transfer from BOS to STO, producing a
$n$-type, BOS/STO-mixed metallic state at the interface. In contrast,
the \mbox{[Se-SrO]} contact type, which is the most stable contact
configuration between BOS and SrO-terminated STO substrate, belongs to
$p$-type and has a much smaller interfacial charge transfer from STO
to BOS, as well as much weaker interfacial electronic
hybridization. These results suggest that the BOS/STO interfaces share many
similarities with the LAO/STO interfaces in terms of interfacial
electronic structure~\cite{Ohtomo2004}, with important implications
for exploring emergent phenomena at the interfaces of oxychalcogenides
and oxides, as well as for optimizing the epitaxial growth of BOS thin
films on oxide substrates for practical device applications.

\section{Methods}
For our calculations we adopted density functional theory (DFT) within
the generalized gradient approximation (GGA) of the
exchange-correlation potential, as parametrized by
Perdew-Burke-Ernzerhof (PBE)~\cite{Perdew1996}. The interaction
between valence electrons and ionic cores was treated using projector
augmented wave (PAW) potentials as implemented in the Vienna Ab initio
Simulation Package (VASP)~\cite{Bloechl1994,Kresse1996}.

To study the properties of ideal BOS/STO interface, we
constructed a heterostructural model consisting of BOS and STO slabs
along each of their [001] directions. The tetragonal supercell of the
slab model corresponds to an in-plane repeating unit of $1 \times 1$
for both BOS and STO, and the in-plane lattice constant of the
supercell follows that of STO, which simulates the coherent epitaxial
growth of ultrathin BOS films on thick STO substrates~\cite{Tan2019,
  Liang2019}. Regarding the specific atomistic arrangement at the
interface, we considered all possible interfacial contact
configurations and a series of corresponding slab models were
constructed and simulated, in order to determine the lowest-energy
configurations. The STO layers in the slab model is always
stoichiometric. Hence, depending on the interfacial contact
configurations, the surface terminations of STO slabs vary on the
vacuum side. The surface of the BOS slab on the vacuum side is
terminated by a layer of Se atoms passivated by hydrogens. The reason
we use this configuration is that real BOS (001) surface usually
consists of a half-full, dimerized Se layer with 50\% of
vacancies~\cite{Chen2018}. We find that hydrogen passivation of a full
Se surface layer can effectively model the electronic properties of a
BOS slab with 50\% of surface Se vacancies (see \textbf{Fig.~S1} in
the Supplemental Materials), while circumventing the need to enlarge 
the in-plane unit cells of the slab, which significantly reduces the
cost of computing the properties of BOS/STO interface.  The chemical
formulas corresponding to the slab models are
(Bi$_2$O$_2$Se)$_4$H/(SrTiO$_3$)$_3$ and
(Bi$_2$O$_2$Se)$_3$SeH/(SrTiO$_3$)$_3$ for interfaces with a
Bi$_2$O$_2$ contact layer and Se contact layer, respectively. The slab
thickness was verified to be sufficient for modelling the electronic
properties of the BOS/STO interfaces (see \textbf{Fig.~S2} in the
Supplemental Materials).

A sufficiently large vacuum thickness of 20~\AA\ was set in the slab
model. Additional dipole correction~\cite{Bengtsson99} on the
electrostatic potential along the vertical direction of the slab was
found to have a negligible influence on the calculated properties. For
the plane-wave expansion of the electronic wavefunctions, a kinetic
energy cutoff of 500~eV is used. The Brillouin zone was sampled using a
9$\times$9$\times$1 Monkhorst-Pack $\mathbf{k}$-point
mesh~\cite{Monkhorst1976}. The convergence threshold for the
self-consistency of the total energy is set to 10$^{-4}$~eV, and all
atoms in the slab were fully relaxed until the force on each atom is
smaller than 0.01~eV\,\AA$^{-1}$.

\begin{figure*}[!t]
	\centering \includegraphics[width=18 cm]{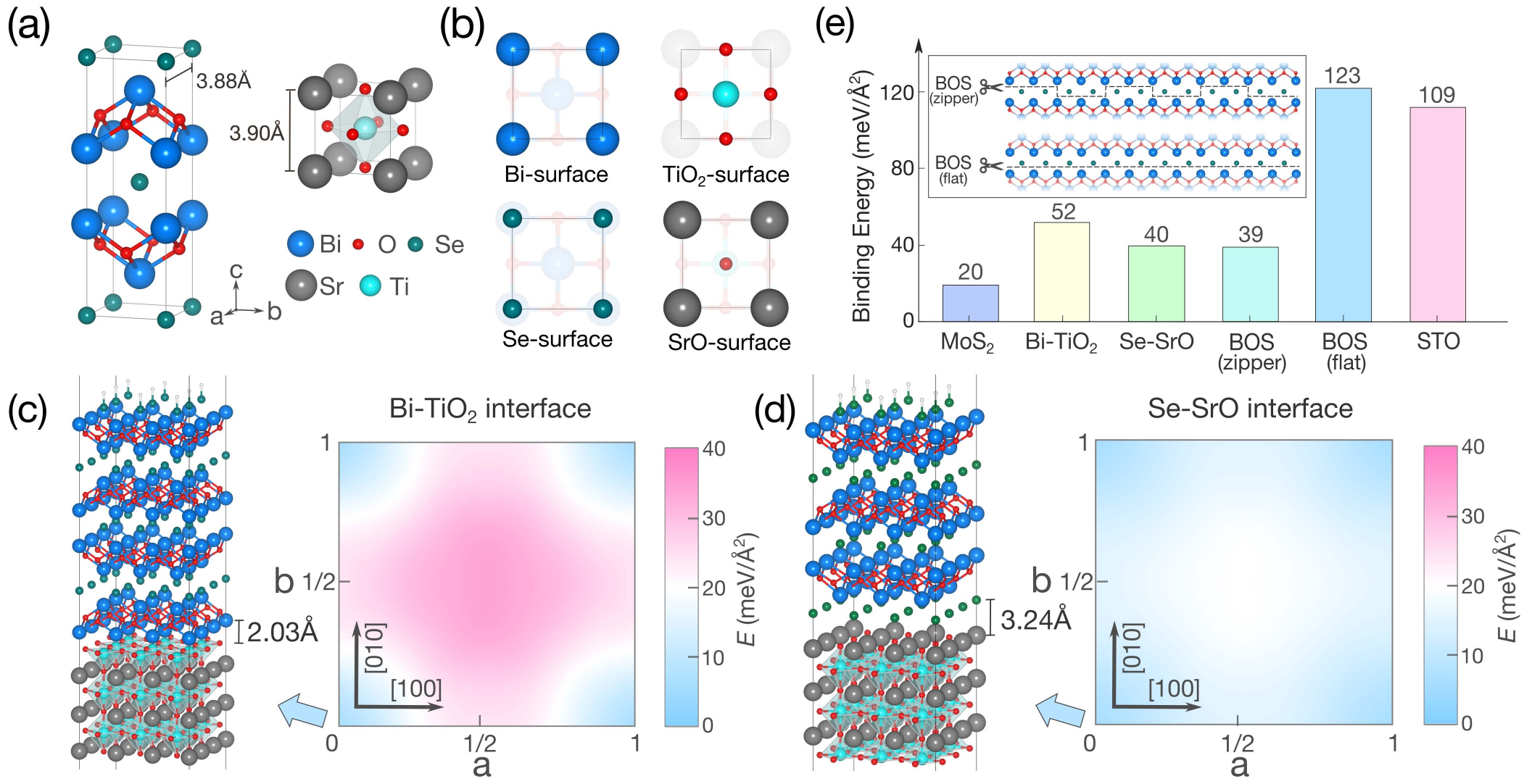}
	\caption{(\textbf{a}) Atomistic models of the conventional
          tetragonal unit cell of \BOS\ (BOS) and the cubic unit cell
          of \STO\ (STO).  (\textbf{b}) Illustrations of the Bi- and
          Se-terminated surfaces of BOS (top- and bottom-left panels),
          as well as the TiO$_2$- and SrO-terminated surfaces of STO
          (top- and bottom-right panels). The representations of atoms
          are the same as those in (\textbf{a}).
          (\textbf{c},\textbf{d}) DFT-relaxed atomistic structural
          models of the lowest-energy [Bi-TiO$_2$] interface
          (\textbf{c}) and [Se-SrO] interface (\textbf{d}). The
          distance between the Bi layer and TiO$_2$ layer at the
          [Bi-TiO$_2$] interface is $\sim$2.0~\AA, whereas the
          distance between the Se layer and SrO layer at the [Se-SrO]
          interface is $\sim$3.2~\AA.  Also shown together are the
          corresponding changes of the interfacial binding energies
          with respective to the rigid relative in-plane displacements
          of BOS and STO slabs along the $a$ and $b$ axes ([100] and
          [010] directions). (\textbf{e}) Interlayer/interface binding
          energies of various systems calculated by density functional
          theory (DFT). The plot includes data for interlayer binding
          energy in MoS$_2$, interface binding energies between BOS
          and STO that belong to the [Bi-TiO$_2$] and [Se-SrO] contact
          types, interlayer binding energies in BOS that correspond to
          zipper or flat cleavage, as well as the interlayer binding
          energy between the TiO$_2$ and SrO sheets in STO. The inset
          illustrates the contact breaking patterns of the zipper and
          flat cleavage of BOS. }
	\label{Fig1}
\end{figure*}

\section{Results}
\subsection{Structure and energetics of the BOS/STO interfaces}
As our goal is to understand the interfacial properties of BOS/STO, we
begin with a discussion of the crystal structures of both BOS and
STO. BOS has a body-centered tetragonal structure with a 4-fold
symmetry ($I4/mmm$ space group), with the experimental lattice
parameters $a=3.88$~\AA, and $c=12.16$~\AA~\cite{Boller1973}. The
structure belongs to the \textit{anti}-ThCr$_2$Si$_2$
type~\cite{Hoffmann1985}, and the corresponding atomistic model is
illustrated in \textbf{Fig.~1(a)} (left panel). In BOS, Bi and O atoms
form layered, covalently bonded frameworks with edge-sharing BiO$_4$
square-pyramid coordination~\cite{Zhu2022,Hoffmann1985}. Between
Bi$_2$O$_2$ layers are Se atoms arranged in a 2D square lattice. As Se
is more electronegative than Bi, the Bi atoms in the Bi$_2$O$_2$
layers transfer electrons to the Se layers, resulting in positively
charged [Bi$_2$O$_2$]$_n^{2n+}$ layers and negatively charged
[Se]$_n^{2n-}$ layers, where $n$ is the number of repetitive in-plane
formula units. The [Bi$_2$O$_2$]$_n^{2n+}$ and [Se]$_n^{2n-}$ layers
are bound together by forces mostly of electrostatic nature. Since the
nominal charge state of Bi in BOS is $+3$, each Bi atom has a lone
pair of 6$s^2$ electrons. The lone pair electrons are stereoactive and
direct in the layer-normal direction, playing a key role in stabilizing
the layered structure~\cite{Pereira2018}.

The natural cleavage plane of BOS is the (001) plane. The cleavage
process leaves 50 percent of Se on each of the two resulting surfaces,
where most of the Se atoms left on the surfaces have a dimerized
structure~\cite{Chen2018}. The interlayer binding energy corresponding
to this dimerized ``zipper'' cleavage mode~\cite{Wei2019} has a small
value of $\sim$39 meV/\AA$^2$, as obtained from our DFT
calculations. By contrast, the DFT-calculated binding energy between
atomically flat Bi$_2$O$_2$ and Se layers, which corresponds to all the Se
atoms on one side of the two surfaces created by cleavage, has a much
higher value of $\sim$123 meV/\AA$^2$. To put these numbers in
perspective, the interlayer binding energy of MoS$_2$ is merely
$\sim$20 meV/\AA$^2$~\cite{Bjorkman2012}. The distinct bonding
characteristics of BOS suggest that its interfacial properties could
be rather different from those of conventional layered semiconductors.

On the other hand, STO has a typical cubic perovskite structure with
$Pm\bar{3}m$ space group and a lattice constant of
3.90~\AA~\cite{Abramov1995}, which is also illustrated in the right
panel of \textbf{Fig.~1(a)}. In the [001] direction, STO has alternate
stacking of planar TiO$_2$ and SrO atomic sheets. Thus, the (001)
surface of STO has two distinct terminations, either TiO$_2$- or
SrO-terminated. As the nominal charge of Sr, Ti, and O in STO are
$+2$, $+4$, and $-2$, respectively, in the simple ionic limit, each
TiO$_2$ or SrO sheet can be considered charge
neutral~\cite{Hwang2012}. Normally, the surface of STO substrates
obtained by cleavage or cutting consists of an equal amount of
TiO$_2$- and SrO-terminated domains separated by half-unit-cell
steps~\cite{Huijben2009}. However, simple chemical treatment methods
have been developed to achieve fully TiO$_2$-terminated
surfaces~\cite{Kawasaki1994, Koster1998, Ohnishi2004}. The opposite
single-terminated SrO surfaces can be obtained either by annealing STO
substrates in air at high temperatures~\cite{Bachelet2009}, or by
depositing a SrO monolayer on top of a single-terminated TiO$_2$
surface~\cite{Hwang2012, Radovic2009}.

The (001) planes of BOS and STO are symmetry matched (both have 2D
square lattices), with a small lattice-constant difference of 0.5\%
(3.88~\AA\ versus 3.90~\AA). Therefore, when a BOS thin film is grown
on the (001) surface of a STO substrate, coherent interface can form,
wherein the in-plane lattice constants of BOS follow those of
STO~\cite{Tan2019, Liang2019}. Since the (001) surface of STO can be
TiO$_2$- or SrO-terminated, whereas that of BOS can be terminated with
a Bi$_2$O$_2$ or Se layer, as illustrated in \textbf{Fig.~1(b)}, four
atomically sharp interfacial contact types are theoretically possible
between BOS and STO. The [Bi-TiO$_2$] contact type involves the direct
contact of a Bi$_2$O$_2$ layer of BOS on top of a TiO$_2$-terminated
STO. This contact type has been experimentally observed in BOS thin
films grown on STO by both CVD and molecular beam
epitaxy~\cite{Tan2019, Liang2019}. The [Se-SrO] contact type
corresponds to a Se layer of BOS on SrO-terminated STO. The
[Se-TiO$_2$] and [Bi-SrO] contact types have similar connotations.

In each of the four contact types, additional in-plane translational
degrees of freedom exist. We have considered all the possible
high-symmetry, coherent interfacial contact configurations between the
(001) surfaces of BOS and STO, and carried out DFT calculations of the
corresponding interfacial binding energies. Here, the interfacial
binding energy is defined as the energy needed to separate an
interface. The results of our calculations indicate that the
[Bi-TiO$_2$] contact type, in its most stable configuration, has the
highest binding energy of 52 ~meV/\AA$^2$ among all four interfacial
contact types. The binding energy of the [Se-SrO] interface has a
close value of 40~meV/\AA$^2$.  In contrast, the other two interfacial
contact types, namely [Se-TiO$_2$] and [Bi-SrO], have much smaller
binding energies of 16 meV/\AA$^2$ and 5 meV/\AA$^2$,
respectively. Hence, when BOS is grown on a TiO$_2$-terminated STO
substrate, the equilibrium interfacial contact type should be
[Bi-TiO$_2$], which is in consistent with previous experimental
observations~\cite{Tan2019, Liang2019}. On the other hand, if the STO
substrate is SrO-terminated, the [Se-SrO] contact type should be
energetically much more competitive than the [Bi-SrO] type. The
energetic order between different contact types can be understood in
terms of their interfacial charge transfer properties, which will be
discussed latter.

The lowest-energy in-plane alignments between BOS and STO in the
[Bi-TiO$_2$] and [Se-SrO] contact types are in accordance with the
illustrated Bi- and Se-surfaces of BOS, as well as the TiO$_2$- and
SrO-surfaces of STO in {\textbf{Fig.~1(b)}}. Specifically, in the
[Bi-TiO$_2$] contact type, the Bi atoms of BOS sit above the four-fold
hollow sites of O atoms and align with the Sr atoms below, in
agreement with experimental observations~\cite{Tan2019, Liang2019}. On
the other hand, in the [Se-SrO] contact type, Se atoms are located
directly on top of Sr atoms.  Additional side views of the
[Bi-TiO$_2$] and [Se-SrO] contact types are shown in
{\textbf{Fig.~1(c,d)}} and Supplementary \textbf{Fig.~S3}.

In \textbf{Fig.~1(e)} we compare the binding energies of
\mbox{[Bi-TiO$_2$]} and [Se-SrO] interfaces with the interfacial or
interlayer binding energies in other systems, including the vdW
interlayer binding energy of MoS$_2$, the interlayer binding energy of
BOS that corresponds to zipper or atomically flat cleavage, as well as
the binding energy between TiO$_2$ and SrO sheets in STO. It is noted
that in the flat cleavage mode of BOS, which leaves all Se atoms on
one side of the cleaved surface, the corresponding binding energy (123
meV/\AA$^2$) is even larger than that of between TiO$_2$ and SrO
sheets in STO (109 meV/\AA$^2$).  This shows that the interlayer
electrostatic interaction between [Bi$_2$O$_2$]$_n^{2n+}$ and
[Se]$_n^{2n-}$ layers in BOS is by no means weak. Furthermore, the
binding energies of the [Bi-TiO$_2$] and [Se-SrO] interfaces are both
several times higher than that between MoS$_2$ layers.  The relatively
strong interfacial binding between BOS and STO, in combination with
symmetry and close lattice-constant matching, makes coherent epitaxial
growth of BOS on STO and other perovskite-related materials growth
possible~\cite{Tan2019, Liang2019}. This creates an exciting
opportunity to generate a large variety of novel heterostructures
between the oxychalcogenide BOS and perovskite oxides.

\begin{figure*}[!t]
	\centering
	\includegraphics[width=0.8\textwidth]{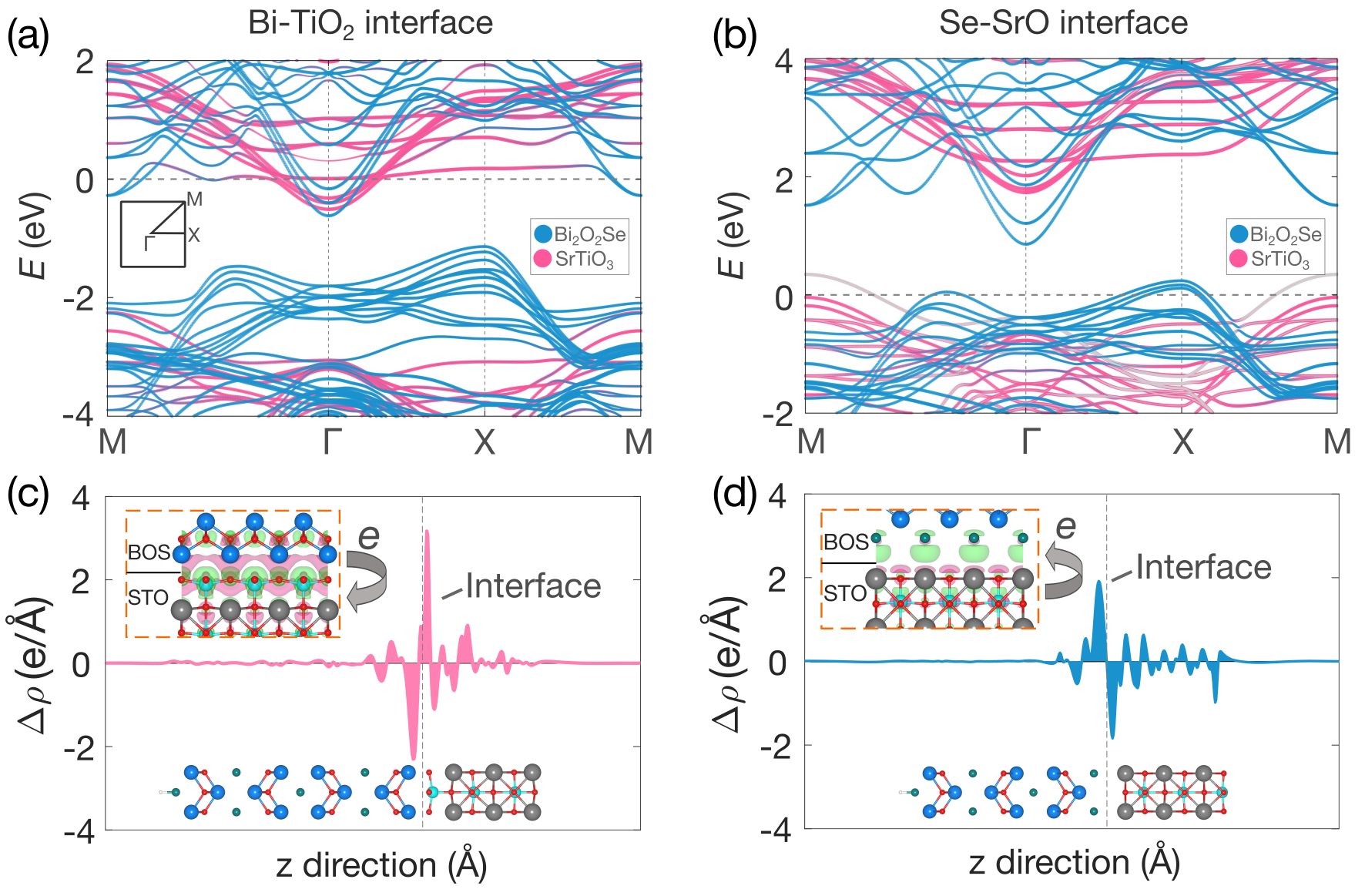}
	\caption{(\textbf{a,b}) Density functional theory (DFT)
          calculated electronic band structures of ideal BOS/STO
          interface with [Bi-TiO$_2$] and [Se-SrO] contact types. The
          inset in (\textbf{a}) is the corresponding reciprocal space
          path. The Fermi level is set to energy
          zero. The gray lines in (\textbf{b}) denote that the
          corresponding electronic states derive from the surface
          states of the STO slab on the opposite side of the interface.
          (\textbf{c,d}) Planar averaged electron density
          differences of the [Bi-TiO$_2$] (\textbf{c}) and [Se-SrO]
          (\textbf{d}) contact types with respect to isolated slabs,
          which are calculated as $\Delta\rho = \rho_{\text{BOS/STO}}
          - \rho_{\text{BOS}} - \rho_{\text{STO}}$. The inset figures
          in dashed boxes are plots of charge density differences,
          drawn using the VESTA software~\cite{Momma2011}. The green
          (pink) color corresponds to charge accumulation (depletion)
          and the isosurfaces correspond to $\Delta\rho$ equal to
          0.0015~e/\AA$^3$.}
	\label{Fig2}
\end{figure*}

\subsection{Electronic properties of the ideal BOS/STO interfaces}
Having studied the contact configurations and energetics of the
BOS/STO interfaces, we next investigate their electronic
properties. \textbf{Fig.~2(a,b)} show the DFT-computed electronic band
structures of the heterostructural slab models of the [Bi-TiO$_2$] and
[Se-SrO] interfaces, which correspond to the atomistic structural
models in Fig.~1(d) and 1(e), respectively. In the case of the
[Bi-TiO$_2$] interface, the Fermi level ($E_\textrm{F}$) is located
above the CBM and crosses both the BOS and STO components, resulting
in a $n$-type metallic phase. Notably, there is a distinct
hybridization of electronic states between BOS and STO near the
$E_\textrm{F}$ along the $\textrm{M}$--$\Gamma$ and
$\textrm{X}$--$\textrm{M}$ directions in its 2D Brillouin zone. On the
other hand, for the [Se-SrO] contact interface, the $E_\textrm{F}$
shifts to the valence band, resulting in a $p$-type metallic
state. However, no noticeable hybridization between BOS and STO can be
observed from the projected band structure in this contact type.

To probe the origin of the metallic states, we compute the electronic
band structures of isolated slabs of BOS and STO. \textbf{Fig.~S4}
shows that the Fermi levels of STO slabs are always located in the
band gap regardless of the surface termination, corresponding to an
insulating state. In contrast, in a Bi$_2$O$_2$-terminated BOS slab, the
Fermi levels move into the conduction band, while in a Se-terminated
slab, the Fermi level is in the valence band. When a BOS slab and a STO
slab are put into contact, charge transfer can occur between BOS and
STO slabs in order to reach a uniform electron chemical potential
across the interface. In the case of [Bi-TiO$_2$] interface, after
electronic reconstruction, the Fermi level crosses the conduction
bands of both BOS and STO, indicating charge transfer from the BOS
slab to the STO slab, while in the case of [Se-SrO] interface, the
Fermi level crosses the valence bands of both BOS and STO, indicating
a reverse direction of charge transfer from STO to BOS.

The different charge transfer behavior is confirmed by explicitly
computing the amount of interfacial charge transfer at the two types
of BOS/STO interfaces. As shown in \textbf{Fig.~2(c,d)}, charge
accumulation and depletion on the side of STO are found in the
[Bi-TiO$_2$] and [Se-SrO] contact types, respectively. In addition to
the opposite charge transfer direction, the magnitude of charge
transfer is significantly different between the two. Bader charge
analysis~\cite{Henkelman2006} reveals that about 0.5 electron per 2D
unit cell ($e$/u.c) is transferred from BOS to STO in the
[Bi-TiO$_2$] contact type, whereas the corresponding number is
0.2~$e$/u.c from STO to BOS in the [Se-SrO] contact type.  The
stronger interfacial charge transfer and electronic hybridization
observed at the [Bi-TiO$_2$] contact interface explain its
higher interfacial binding energy than the [Se-SrO] contact type, as
discussed in the previous section.

\section{Discussions}
The intriguing electronic properties of the BOS/STO interfaces
originate from polar discontinuity at the contact interfaces.  Unlike
vdW layered materials such as MoS$_2$, BOS consists of alternate
stacking of the charged [Bi$_2$O$_2$]$^{2+}$ and [Se]$^{2-}$ layers
along the [001] direction. Each Bi$_2$O$_2$ layer donates two
electrons per 2D unit cell to the neighboring Se layers, resulting in
interlayer electrostatic interaction between the [Bi$_2$O$_2$]$^{2+}$
and [Se]$^{2-}$ layers. Here, the nominal charge of Bi, O, and Se are
$3+$, $2-$, and $2-$ in the simple ionic limit, respectively. In
comparison, STO has alternate stacking of charge neutral TiO$_2$ and
SrO layers along the [001] direction, with the nominal charges of Ti,
Sr, O being $4+$, $2+$, and $2-$ in the simple ionic limit,
respectively. Hence, along the [001] direction, BOS is polar while STO
is not, and as a result, there is a polar discontinuity at the
interface between BOS and STO, similar to that at the LAO/STO
interface~\cite{Ohtomo2004,Mannhart2008,Hwang2012}. For the
[Bi-TiO$_2$] contact type of the BOS/STO interface, the polar
discontinuity can be denoted as (Bi$_2$O$_2$)$^{2+}$/(TiO$_2$)$^0$,
while for the [Se-SrO] contact type, it can be denoted as
(Se)$^{2-}$/(SrO)$^0$.

In the absence of interfacial electronic or atomic reconstruction,
polar discontinuity at heterostructural interface can lead to ``polar
catastrophe'', where the electrostatic potential generated by the
charged layers grows quickly away from the interface and diverges in
the bulk limit~\cite{Nakagawa2006, Hwang2006, Mannhart2008}. In
heterostructures formed by growing a conventional polar semiconductor
on a non-polar semiconductor, such as GaAs on Si, such potential
divergence is avoided via atomic reconstruction at the interface
through atomic disordering or a change in
stoichiometry~\cite{Baraff1977,Harrison1978,Kroemer1987}. However, in
oxide heterostructures where certain ions can assume multiple
valencies, such as Ti in STO, whose valence can change from Ti$^{4+}$
to Ti$^{3+}$, polar catastrophe can be avoided via electronic
reconstruction, specifically through interfacial charge transfer that
leads to mixed valencies of certain ions~\cite{Nakagawa2006,
  Hwang2006, Mannhart2008}. Such interfacial electronic construction
was considered to be responsible for the observation of 2D electron
gas at the (LaO)$^{+}$/(TiO$_2$)$^0$ interface between LAO and
STO~\cite{Ohtomo2004}.

The interfacial electronic properties of the BOS/STO interfaces thus
share significant similarities with the LAO/STO interfaces. The polar
discontinuity of the \mbox{[Bi-TiO$_2$]} contact interface, explicitly
written as (Bi$_2$O$_2$)$^{2+}$/(TiO$_2$)$^0$, corresponds to that of
the (LaO)$^{+}$/(TiO$_2$)$^0$ interface between LAO and STO. After
electronic reconstruction in self-consistent DFT calculations, this
interface type involves charge transfer from the Bi$_2$O$_2$ layer to
the TiO$_2$ layer in the amount of ${\sim}0.5$~$e$/u.c., as
illustrated in \textbf{Fig.~3(a)}. Conceptually, one can also start
from the atomic limit and then allow ionization of the elements. In
bulk BOS, a Bi$_2$O$_2$ layer donates 2~$e$/u.c. of electrons to its two
neighboring Se layers. However, for the Bi$_2$O$_2$ layer at the
Bi$_2$O$_2$/TiO$_2$ interface, only one Se layer on one side of the
interface is available to accept electrons. Hence, the Bi$_2$O$_2$
layer at the interface have excess electrons on the order of
1~$e$/u.c, which leads to a $n$-type interface. These excess electrons can
be partially transferred to the TiO$_2$ layer on the other side of the
interface, in the process partially changing the Ti valency from
Ti$^{4+}$ to Ti$^{3+}$.

\begin{figure}[t!]
	\centering
	\includegraphics[width=0.45\textwidth]{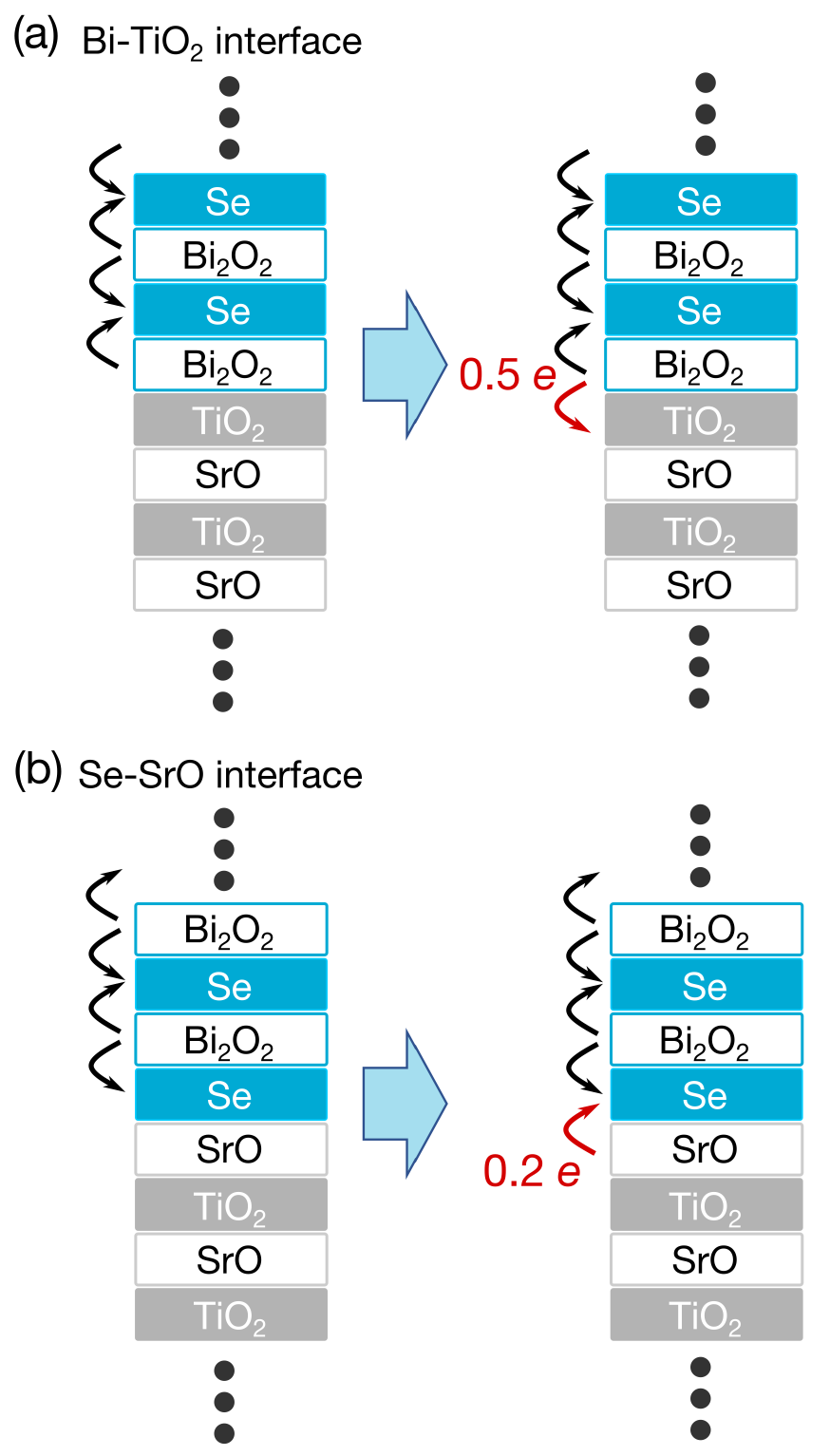}
	\caption{Schematic diagrams of polar discontinuities and
          charge transfer at the interfaces between BOS and
          STO. (\textbf{a}) The [Bi-TiO$_2$] contact
          interface. (\textbf{b}) The [Se-SrO] contact interface.}
	\label{Fig3}
\end{figure}

On the other hand, the polar discontinuity at the \mbox{[Se-SrO]} contact
interface, written as (Se)$^{2-}$/(SrO)$^0$, is akin to the
(AlO$_2$)$^{-}$/(SrO)$^0$ interface between LAO and STO. In this case,
the Se layer at the interface only has one neighboring Bi$_2$O$_2$
layer from which it can receive electrons, leading to a $p$-type
electronic state. This electron deficiency could be mitigated by
charge transfer from the SrO layer on the STO side of the
interface. However, since it is energetically costly to induce mixed
valency for either Sr or O, the amount of charge transfer from the SrO
layer is rather limited. In our DFT calculation of the ideal \mbox{[Se-SrO]}
interface, the amount of charge transfer is calculated to be
${\sim}0.2$~$e$/u.c., as illustrated in \textbf{Fig.~3(b)}. In fact, we find
that the transferred charge in our DFT calculation mainly originates
from the surface state of the STO slab on the opposite side of the
contact interface. Given that in real
experiments, the SrO-terminated substrate is rather thick, charge
transfer from the STO side to the BOS side would be further
hampered. Since electronic reconstruction at the \mbox{[Se-SrO]} contact
interface is energetically unfavorable, atomic reconstruction is expected
to occur in this contact type in real systems, where the generation
of Se vacancies or O vacancies near the interface can provide the extra
electrons that compensate the $p$-type carriers. Indeed, in the
$p$-type (AlO$_2$)$^{-}$/(SrO)$^{0}$ interface between LAO and STO, oxygen vacancies
play a central role in avoiding polar catastrophe and lead to an
insulating interface instead of $p$-type hole
transport~\cite{Ohtomo2004, Nakagawa2006}. Similar phenomena could
happen at the \mbox{[Se-SrO]} contact type of the BOS/STO interface.

It is worth emphasizing that the polar discontinuities of the BOS/STO
interfaces arise from the charged layered structure of BOS with
interlayer electrostatic interaction, which is absent in
heterostructures of typical vdW layered materials such as MoS$_2$
grown on STO. Consequently, the substrate used for growing BOS thin films
could have a significant influence on their experimentally measured
electronic properties. When mica substrates are used for the CVD
growth of BOS~\cite{Wu2017n}, the interfacial structure involves the
surface K$^+$ layer of mica in contact with the Se layer of BOS, as
illustrated in \textbf{Fig.~4(a)}, which has been experimentally
observed using atomic-scale scanning transmission electron microscopy
(STEM)~\cite{Hong2020}. Since the K$^+$ layer of mica is
electronically inert, charge transfer between BOS and mica shall be
minimal. By contrast, for BOS thin films grown on TiO$_2$-terminated
STO substrates, the interface belongs to the \mbox{[Bi-TiO$_2$]}
contact type~\cite{Tan2019, Liang2019}, as schematically depicted in
\textbf{Fig.~4(b)}. As we have discussed above, due to polar discontinuity
and the ability of Ti$^{4+}$ to undergo valence change,
significant interfacial charge transfer from BOS to STO is expected to
occur. Consequently, the electronic states responsible for the
$n$-type electron transport have contributions from both BOS and STO.
As the electron effective mass of STO is larger than that of BOS (see
also Fig.~2(a)), the resulting measured room-temperature electron
mobility of BOS grown on TiO$_2$-terminated STO substrates shall be
smaller than those measured from BOS samples grown on mica substrates,
which was indeed found in experiments~\cite{Tan2019}. Nonetheless, the
strong interfacial interaction means that novel electronic or magnetic
properties could emerge from the \mbox{[Bi-TiO$_2$]}
contact type of the BOS/STO interface, as in the case of LAO grown on
TiO$_2$-terminated STO substrates~\cite{Hwang2012}.

For the \mbox{[Se-SrO]} contact configuration, the interfacial charge
transfer is much weaker, resulting in a weakly $p$-type electronic
structure and energetically well-separated conduction bands of BOS and
STO (Fig.~2(b)). Although the formation of Se vacancies and possibly
oxygen vacancies, promoted by the polar discontinuity at the
interface, are likely to compensate the hole carriers, the
$p$-type behavior before reconstruction suggests
that field-effect transistors fabricated from BOS layers grown on
SrO-terminated STO substrate, as illustrated in \textbf{Fig.~4(c)},
may circumvent the problem of high residual electron carrier
concentrations that has plagued BOS-based transistor
devices~\cite{Wang2022}, and may even pave the way for realizing
$p$-type transistors based on BOS.

\begin{figure}[t]
	\centering
	\includegraphics[width=0.48\textwidth]{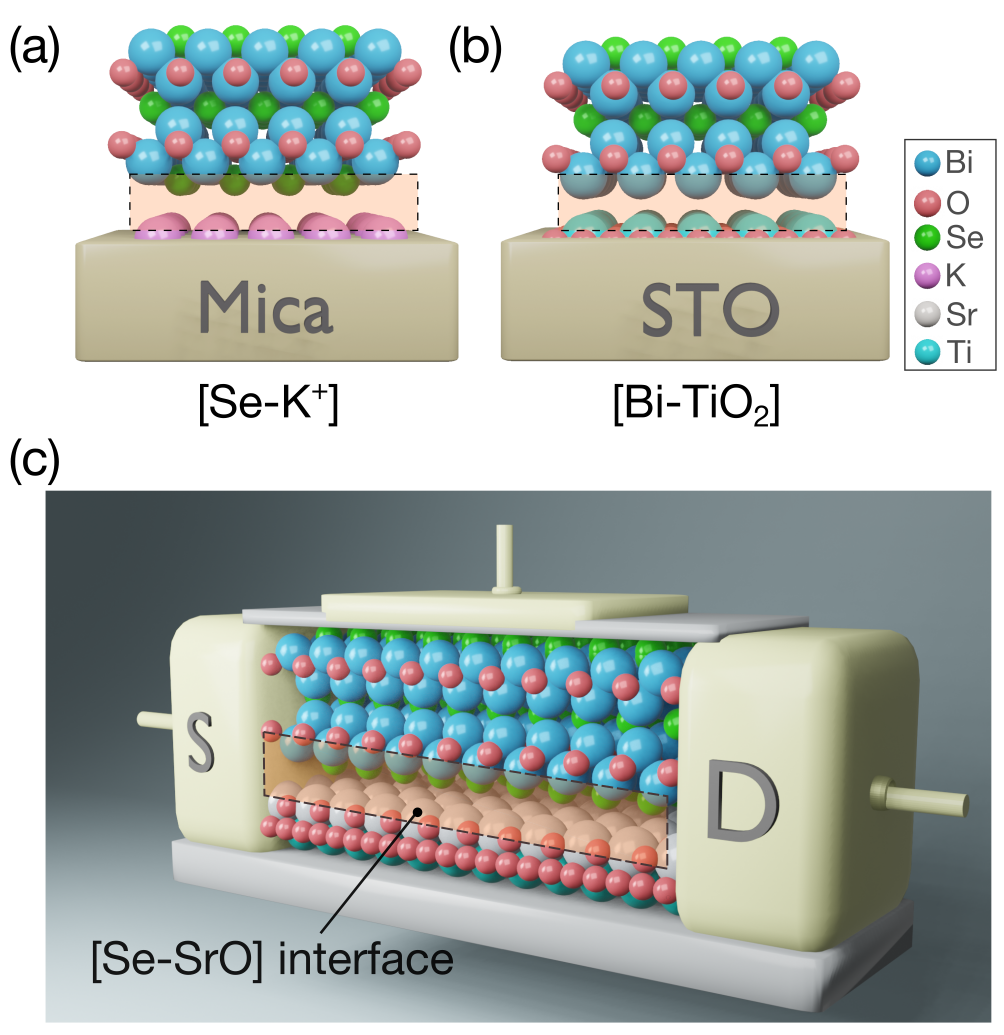}
	\caption{(\textbf{a},\textbf{b}) Illustration of the contact
          interface of BOS thin films grown on a mica and
          TiO$_2$-terminated STO substrates. (\textbf{c}) Schematic
          illustration of a field-effect transistor (FET) device
          directly fabricated on a BOS layer grown on a
          SrO-terminated STO substrate, with the \mbox{[Se-SrO]}
          contact interface indicated by orange shade.}
	\label{Fig4}
\end{figure}

\section{Conclusions}
In conclusion, we have performed systematic first-principles
investigations of the energetic and electronic properties of ideal
BOS/STO interfaces, revealing the crucial effect of interfacial
contact type and polar discontinuity on the interfacial
properties. For BOS grown on TiO$_2$-terminated STO, a Bi$_2$O$_2$
layer in contact with a surface TiO$_2$ layer of STO, denoted by
\mbox{[Bi-TiO$_2$]}, is found to be the lowest-energy interface
configuration, in consistent with experimental observations. Due to
the polar discontinuity between BOS and STO and the ability of
titanium ions to exist in mixed valencies, a significant charge
transfer from BOS to STO occurs in the \mbox{[Bi-TiO$_2$]} contact
type. This leads to a $n$-type interface with a mixed metallic state
of BOS and STO, which is likely responsible for the lower electron
mobility observed in BOS grown on TiO$_2$-terminated STO substrates as
compared to BOS grown on electrically more inert substrates such as
mica. For SrO-terminated STO substrates, we find that the
lowest-energy contact configuration between BOS and STO is a Se layer
on a SrO layer, denoted by \mbox{[Se-SrO]}. This contact configuration
has a much weaker interfacial charge transfer, resulting in a $p$-type
electronic structure. The asymmetry in interfacial charge transfer
properties between \mbox{[Bi-TiO$_2$]} and \mbox{[Se-SrO]} interfaces
can be explained in terms of the ability of Ti$^{4+}$ to undergo a
change in valence to Ti$^{3+}$ in a TiO$_2$ layer, whereas such a
change in valence is energetically costly for Sr$^{2+}$ or O$^{2-}$ in
a SrO layer. In real experiments, the hole carriers at the
\mbox{[Se-SrO]} interface may be compensated by ionized Se or oxygen
vacancies. However, the energetically well-separated conduction bands
of BOS and STO in this contact type suggests that the excellent
electron transport properties of BOS may be better preserved in this
contact type. These results indicate that BOS grown on
TiO$_2$-terminated STO substrates could be a fruitful system for
exploring emergent interface phenomena between an oxychalcogenide and
an oxide, whereas SrO-terminated STO substrates may be desirable for
wafer-scale growth of BOS films with excellent carrier transport
properties.

\section*{Acknowledgments}
This work is supported by NSFC under Project No. 62004172. W.L. also
acknowledges the support by Research Center for Industries of the
Future at Westlake University under Award No. WU2022C041. X.L. and
X.Z. thank the Westlake Multidisciplinary Research Initiative Center
(MRIC) under Grant No. MRIC20200402. The authors
thank Shu Zhao and Xiaoping Yao for helpful discussions, as well as
the High-Performance Computing Center of Westlake University for
technical assistance.


%

\end{document}